\documentclass[prd,nofootinbib,amsfonts,tightenlines,12pt]{revtex4}
\pdfoutput=1

\usepackage{graphicx} 
\usepackage{amsmath}
\usepackage{amssymb}
\usepackage{multirow}
\usepackage{xspace}
\usepackage[utf8]{inputenc}
\usepackage[usenames,dvipsnames]{color}
\usepackage[bookmarks=true,bookmarksnumbered=true]{hyperref}
 \hypersetup{
     colorlinks=true,
     linkcolor=Blue,          
     citecolor=Blue,
     urlcolor=Blue
     } 
     
\usepackage[utf8]{inputenc} 
     
\def\be#1{\begin{equation}\label{#1}}
\newcommand{\ee}{\end{equation}}

\newcommand{\ie}{\textit{i.e.}, }
\newcommand{\eg}{\textit{e.g.}, }
\newcommand{\dnew}{\ensuremath{\delta_{24}}\xspace}

\begin{document}  

\begin{flushright}
FERMILAB-PUB-17-271-T
\end{flushright}

\title{DUNE sensitivities to the mixing between \\ sterile and tau neutrinos}
\author{Pilar Coloma~$^1$}\email{pcoloma@fnal.gov}
\author{David V. Forero~$^{2,3}$}\email{dvanegas@ifi.unicamp.br}
\author{Stephen J. Parke~$^1$}\email{parke@fnal.gov}
\affiliation{$^1$~Theoretical Physics Department, Fermi National Accelerator Laboratory,
P.O. Box 500, Batavia, IL 60510, USA}
\affiliation{$^2$~Instituto de F\'{i}sica Gleb Wataghin - UNICAMP,
13083-859, Campinas, SP, Brazil}
\affiliation{$^3$~Center for Neutrino Physics, Virginia Tech,
  Blacksburg, VA 24061, USA}
\date{July 17, 2017}

\begin{abstract}
Light sterile neutrinos can be probed in a number of ways, including electroweak decays, cosmology and neutrino oscillation experiments. At long-baseline experiments, the neutral-current data is directly sensitive to the presence of light sterile neutrinos: once the active neutrinos have oscillated into a sterile state, a depletion in the neutral-current data sample is expected since they do not interact with the $Z$ boson. This channel offers a direct avenue to probe the mixing between a sterile neutrino and the tau neutrino, which remains largely unconstrained by current data. In this work, we study the potential of the DUNE experiment to constrain the mixing angle which parametrizes this mixing, $\theta_{34}$, through the observation of neutral-current events at the far detector. We find that DUNE will be able to improve significantly over current constraints thanks to its large statistics and excellent discrimination between neutral- and charged-current events. 
\end{abstract}

\maketitle

\newpage

\section{Introduction}

In the past decade, a tremendous experimental effort has been carried out in order to constrain scenarios with additional neutrinos with masses below the electroweak scale. LEP data places severe constraints on the invisible decay of the $Z$. Hence, if there are additional neutrinos below the electroweak scale, they cannot couple to the Standard Model weak bosons (\ie they should be \emph{sterile}). Light sterile neutrinos can lead to observable phenomena in a number of electroweak processes through their impact on the unitarity of the leptonic mixing matrix, including meson decays, muon decay, neutrinoless double beta decay and charged lepton flavor violating transitions (see \eg Refs.~\cite{Antusch:2014woa,Fernandez-Martinez:2016lgt} for recent global fits using these observables). Nevertheless, if their masses are light enough so that they are kinematically accessible in these processes, unitarity is effectively restored at low energies and the bounds from electroweak processes fade away. In this case the best limits are derived from oscillation data~\cite{Declais:1994su,Abe:2014gda,MINOS:2016viw,Astier:2003gs,Astier:2001yj,Parke:2015goa}, see \eg Refs.~\cite{Blennow:2016jkn, Escrihuela:2016ube} for a detailed discussion of these constraints.

In recent years, the eV-scale has recently been put on the spot due to a set of experimental anomalies independently reported in LSND~\cite{Aguilar:2001ty}, MiniBooNE~\cite{AguilarArevalo:2007it,AguilarArevalo:2010wv}, reactor~\cite{Mueller:2011nm, Huber:2011wv} and Gallium experiments~\cite{Giunti:2010zu}. The current and next generation of oscillation experiments will attempt to refute or confirm these hints. The Icecube experiment has recently put impressive limits on the mixing between sterile neutrinos and muon neutrinos $U_{\mu 4}$~\cite{TheIceCube:2016oqi,Aartsen:2017bap}, while in the electron sector strong bounds on $U_{e4}$ have been set by the Daya Bay experiment~\cite{An:2016luf}. In the near future, experiments such as SOX~\cite{Borexino:2013xxa} or STEREO~\cite{Helaine:2016bmc} (among others) will constrain further the mixing with electron neutrinos, while the short-baseline neutrino program at Fermilab will tighten the bounds on the mixing with muon neutrinos~\cite{Antonello:2015lea}. A joint analysis of Bugey-3, Daya Bay and MINOS data has also been performed to constrain the cross-product $|U_{e4}|^2 |U_{\mu 4}|^2$~\cite{Adamson:2016jku}. Conversely, placing equally competitive limits on the mixing with tau neutrinos is a much more difficult task, due to the technical challenges associated to the production and detection of a $\nu_\tau$ beam. 

Indirect constraints on the mixing with $\nu_\tau$ can be derived from the observation of matter effects in atmospheric neutrino oscillations. For example, the Super-Kamiokande experiment sets the bound $|U_{\tau 4}|^2 < 0.18 $ (at 90\% CL) for an active-sterile mass splitting above $0.1~\textrm{eV}^2$~\cite{Abe:2014gda}. On the other hand, a more direct test for the mixing between sterile neutrinos and tau neutrinos can be performed using long-baseline experiments. At long-baseline experiments most of the initial $\nu_\mu$ flux has oscillated into tau neutrinos by the time it reaches the far detector, thanks to $\nu_\mu \to \nu_\tau$ oscillations driven by the atmospheric mass-squared splitting. The OPERA experiment has constrained the impact of sterile neutrinos on this oscillation channel, using charged-current $\nu_\tau$ events at the far detector, setting the bound $4|U_{\mu 4}|^2 |U_{\tau 4}|^2 < 0.116$ (at 90\% CL) for an active-sterile mass-squared splitting above $0.1~\textrm{eV}^2$~\cite{Agafonova:2015neo}. However, their results are severely limited by statistics, since the $\nu_\tau$ charged-current cross section is still low at multi-GeV neutrino energies. 

Alternatively, the mixing between sterile neutrinos and tau neutrinos can be tested at long-baseline experiments searching for a depletion in the neutral-current event rates at the far detector. In fact, both the MINOS and the NOvA experiments have provided competitive constraints using this approach~\cite{Adamson:2011ku,Adamson:2017zcg}. Future long-baseline experiments, with larger detectors, more powerful beams and a better control of systematic uncertainties, may be able to push these limits even further. In this work, we focus on the potential of the DUNE experiment~\cite{Acciarri:2015uup}. Previous studies of sterile neutrino oscillations using the DUNE far detector data can be found, \eg in Refs.~\cite{Berryman:2015nua,Gandhi:2015xza,Agarwalla:2016xxa,Blennow:2016jkn,Agarwalla:2016xlg,Dutta:2016glq, Rout:2017udo}. However, to the best of our knowledge the neutral-current data sample has not been considered in any of these works. The liquid Argon detector technology has excellent particle identification capabilities and therefore a very good discrimination power between charged- and neutral-current events. In addition, the statistics collected at DUNE will exceed considerably (by a rough order of magnitude) the number of events collected at MINOS or NOvA. Thus, DUNE offers an excellent benchmark to conduct a search for sterile neutrino mixing using neutral-current data. 

The manuscript is organized as follows. In section~\ref{sec:probabilities} we derive the oscillation probabilities in the $ \nu_\mu \to \nu_s$ and $ \bar\nu_\mu \to\bar\nu_s$ oscillation channels at the far detector of long-baseline experiments, and discuss the different limits of interest depending on the active-sterile mass-squared splitting. Section~\ref{sec:simulation} summarizes the main features of the DUNE experiment and the details relevant to our numerical simulations. Our results are presented in Section~\ref{sec:results}, and in Section~\ref{sec:conclusions} we summarize and draw our conclusions. Some useful expressions for the elements of the mixing matrix using our parametrization can be found in Appendix~\ref{app}.

%
\section{Oscillation probabilities in the $3+1$ framework}
\label{sec:probabilities}

In this section we derive approximate expressions for the oscillation probabilities, which will be useful in understanding the results of our numerical simulations later on. The mixing matrix $U$ that changes from the flavor to the mass basis in the $3+1$ neutrino framework is a $4\times 4$ unitary matrix: 
\[
\nu_\alpha = U^*_{\alpha i} \nu_i \, ,
\] 
where $\alpha \equiv e,\mu,\tau,s$ and $i \equiv 1,2,3,4$.
In this work we are interested in the effect of oscillations into sterile states on the event rates measured at the DUNE far detector. Assuming that no oscillations have taken place at the near detector, this can be done searching for a depletion in the number of neutral-current (NC) events at the far detector with respect to the prediction obtained using near detector data. For a perfect beam of muon neutrinos with flux $\phi_{\nu_\mu}$ (\ie assuming no beam contamination from other neutrino flavors), the number of NC events at the far detector can be expressed as:
\begin{equation}
\begin{split}
N_{NC} = N_{NC}^e + N_{NC}^\mu + N_{NC}^\tau 
= & \; \phi_{\nu_\mu} \,\sigma_{\nu}^{NC} \left\{ P(\nu_\mu \to \nu_e) + P(\nu_\mu \to \nu_\mu) + P(\nu_\mu \to \nu_\tau) \right\}   \\ 
= & \; \phi_{\nu_\mu} \,\sigma_{\nu}^{NC} \left\{ 1 - P(\nu_\mu \to \nu_s) \right\} \, ,
\end{split}
\end{equation}
and is therefore sensitive to oscillations in the $\nu_\mu \to \nu_s$ channel. Here, $\sigma_\nu^{NC}$ is the neutral-current cross section for the active neutrinos, which is independent of the neutrino flavor. In the absence of a sterile neutrino, the NC event rates should be the same at the far and near detectors up to a known normalization factor coming from the different distance, detector mass, efficiency, and the different geometric acceptance of the beam at the two sites. In fact, the combined fit between near and far detector data should provide a very efficient cancellation of systematic errors associated to the flux and cross section in this channel~\cite{Adamson:2017zcg}.

In addition to the standard solar and atmospheric mass-squared differences, in the $3+1$ framework the oscillation probabilities depend on three new splittings $\Delta m^2_{4k} \equiv m_4^2 - m_k^2$, with $k=1,2,3$. Given the values of the neutrino energy and distance corresponding to the far detector at DUNE, for illustration purposes we can effectively neglect the solar mass splitting and focus on the effects of the oscillation due to the atmospheric and the sterile mass-squared splittings\footnote{In our numerical simulations the full Hamiltonian is diagonalized to extract the oscillation probabilities exactly.}. Under the approximation $\Delta_{21} \ll \Delta_{31}, \Delta_{41}$, the oscillation probability in the $\nu_\mu \to \nu_s$ channel is given (in vacuum) by:
\begin{equation}
\begin{split}
P_{\mu s} \equiv P(\nu_\mu \to \nu_s) &=4 |U_{\mu 4}|^2 |U_{s 4}|^2 \sin^2{\Delta_{41}} + 4 |U_{\mu 3}|^2 |U_{s 3}|^2 \sin^2{\Delta_{31}}  \\
&+ 8 \,\text{Re}\left[U_{\mu 4}^* U_{s 4} U_{\mu 3} U_{s 3}^*\right] \cos{\Delta_{43}}  
 \sin{\Delta_{41}}\sin{\Delta_{31}}  \\
&+ 8 \,\text{Im}\left[U_{\mu 4}^* U_{s 4} U_{\mu 3} U_{s 3}^*\right]\sin{\Delta_{43}}  \sin{\Delta_{41}}\sin{\Delta_{31}},
\end{split}
\label{eq:prob-gral2}
\end{equation}
where we have defined $\Delta_{ij}\equiv \Delta m^2_{ij} L/4 E$. 

The probability in Eq.~\eqref{eq:prob-gral2} is completely general, but does not allow to see the number of independent parameters which enter the oscillation probability. A $4\times 4$ unitary matrix $U$ can be parametrized in terms of six mixing angles and three Dirac CP-violating phases\footnote{If neutrinos are Majorana, additional CP-phases enter the matrix. However, neutrino oscillations are insensitive to these and therefore they will be ignored here. }. In the following, we choose to parametrize it as the product of the following consecutive rotations:
\begin{equation}
U = O_{34} V_{24} V_{14} O_{23} V_{13} O_{12}.
\label{eq:paramet}
\end{equation}
Here, $O_{ij}$ denotes a real rotation with an angle $\theta_{ij}$ affecting the $i$ and $j$ sub-block of the mixing matrix, while $V_{ij}$ denotes a similar rotation but this time including a complex phase. For example:
\begin{equation}
O_{34} = \left( \begin{array}{cccc} 
1 & 0 & 0 & 0 \\
0 & 1 & 0 & 0 \\
0 & 0 & c_{34} & s_{34} \\
0 & 0 & -s_{34} & c_{34}  
\end{array}\right), \quad 
V_{24} = \left( \begin{array}{cccc} 
1 & 0 & 0 & 0 \\
0 & c_{24} & 0 & s_{24} e^{-i \delta_{24}} \\
0 & 0 & 1 & 0 \\
0 & -s_{24}e^{i \delta_{24}} & 0 & c_{24} 
\end{array}\right),
\label{eq:Oij}
\end{equation}
where $s_{ij} \equiv \sin\theta_{ij}$ and $c_{ij} \equiv \cos\theta_{ij}$. In this notation, $ \theta_{i4}$ are the new mixing angles with the fourth state, and $\delta_{14}, \delta_{24}$ are the two new CP-violating phases. In this parametrization, the complex phase associated with the $V_{13}$ rotation corresponds to the standard CP-violating phase in three-families, $\delta_{13} \equiv \delta_{CP}$, and the $3\times 3$ sub-block of the matrix shows only small deviations from a unitary matrix, which at leading order are proportional to $s_{j4}^2$ and therefore within current bounds~\cite{Parke:2015goa}.

For simplicity, from now on we consider $\theta_{14}=0$, which is a valid approximation given the strong constraints set by reactor experiments in the range of $\Delta m^2_{41}$ considered in this work~\cite{Adamson:2016jku}. In this case there is no sensitivity to the $\delta_{14}$ phase, which disappears from the mixing matrix, and the relevant elements of the mixing matrix read
\begin{equation}
\begin{array}{ll}
U_{\mu3} = c_{24} c_{13} s_{23}\, ,    & U_{\mu4} = s_{24}  e^{-i \delta_{24}} \, , \\
U_{s3} = -s_{34}c_{13}c_{23} - s_{24}c_{34}c_{13}s_{23}  e^{i \delta_{24}} \, , & U_{s4}= c_{34}c_{24} \, ,
\end{array}
\end{equation}
see Eq.~(\ref{eq:elements}). Then we can rewrite the $\nu_\mu \rightarrow \nu_s$ oscillation probability, Eq.~\eqref{eq:prob-gral2}, as
\begin{equation}
\begin{split}
P_{\mu s} 
 =  &~c^2_{34} \sin^22\theta_{24}  \, \sin^2 \Delta_{41} \\
&+ 2 c^4_{13} s^2_{23}c^2_{24}\left[
2c^2_{23} s^2_{34} + 
 \sin 2\theta_{23} \sin2\theta_{34} s_{24} \cos\delta_{24} 
 +  2s_{23}^2 c_{34}^2 s_{24}^2 \right]\, \sin^2 \Delta_{31}  \\
&\hspace*{-1cm}  - \left[ c^2_{13}c_{24}  \sin 2\theta_{23} \sin 2\theta_{24}\sin 2\theta_{34} 
\cos(\Delta_{43} - \delta_{24}) +  2 c^2_{13} s^2_{23}  c^2_{34}\sin^22\theta_{24}  \cos\Delta_{43} \right] \sin\Delta_{41} \sin\Delta_{31} ,
\label{eq:prob-final}
\end{split}
\end{equation}
where the dependence with the new CP-violating phase $\delta_{24}$ phase is now evident. 
Depending on the value of the new mass-squared splitting, $\Delta m^2_{41}$, the following three limiting cases can be considered for the probability in Eq.~\eqref{eq:prob-final}:
\begin{enumerate}
\item The oscillations due to the active-sterile mass-squared splitting have not developed at the far detector (\ie $\Delta_{41} \ll \Delta_{31}$):
\begin{equation}
\begin{split}
P_{\mu s} = & ~4 |U_{\mu 3}|^2 |U_{s 3}|^2 \sin^2{\Delta_{31}}  \\[2mm]
= & ~2 c^4_{13} s^2_{23}c^2_{24}\left[
2 c^2_{23} s^2_{34} + 
 \sin 2\theta_{23} \sin 2\theta_{34} s_{24} \cos\delta_{24} + 
 2 s_{23}^2 s_{24}^2 c_{34}^2  \right] \,
\sin^2 \Delta_{31} \, .
\end{split}
\label{eq:dm41to0}
\end{equation}
\item The oscillation maximum due to the active-sterile mass-squared splitting matches the distance to the far detector (\ie $\Delta_{41} \approx \Delta_{31}$):
\begin{equation}
\label{eq:dm41todm31}
\begin{split}
P_{\mu s} = & ~4 \left| U^*_{\mu 4} U_{s 4}+U^*_{\mu 3} U_{s 3} \right|^2  \sin^2\Delta_{31}  \\[2mm]
= & ~4 \left\{ |U_{\mu 4}|^2 |U_{s 4}|^2 + |U_{\mu 3}|^2 |U_{s 3}|^2
+2\, \text{Re}[U_{\mu 4}^* U_{s 4} U_{\mu 3} U_{s 3}^*]\right\}  \sin^2\Delta_{31}  \\[2mm]
 = & ~\left\{  c_{13}^4\sin^22\theta_{23}c_{24}^2 s_{34}^2  + 
c^2_{34} \sin^22\theta_{24} ( 1 - c_{13}^2s_{23}^2)^2    \right. \\ 
 &~ \left.  - c^2_{13} c_{24} \sin 2\theta_{23} \sin 2\theta_{24} \sin 2\theta_{34} (1 - c_{13}^2 s_{23}^2 ) \cos\delta_{24}  \right\} \sin^2\Delta_{31} \, .
\end{split}
\end{equation}
Note that if $U^*_{\mu 4} U_{s 4} + U^*_{\mu 3} U_{s 3} \approx 0 $ there is a significant cancellation in the probability. This will be discussed in more detail later in this section.

\item The oscillations due to the active-sterile mass-splitting are already averaged-out at the far detector\footnote{A similar expression in this limit, but assuming a real mixing matrix, can be found in Ref.~\cite{Adamson:2010wi}.}  (\ie $\Delta_{41} \gg \Delta_{31}$):
\begin{equation}
\begin{split}
P_{\mu s}  = & ~2 \, |U_{\mu 4}|^2 |U_{s 4}|^2 
 + 4 \left\{ |U_{\mu 3}|^2 |U_{s 3}|^2
 + \text{Re}[U_{\mu 4}^* U_{s 4} U_{\mu 3} U_{s 3}^*]\right\} \sin^2 \Delta_{31}  \\[1mm] &
 + 2\, \text{Im}[U_{\mu 4}^* U_{s 4} U_{\mu 3} U_{s 3}^*] \sin 2 \Delta_{31}  \\[2mm]
  =  & ~\frac{1}{2} c^2_{34} \sin^22\theta_{24} \\&
  + \bigg[ 
c_{13}^4\sin^22\theta_{23} c_{24}^2 s_{34}^2  
- c_{13}^2 s_{23}^2 (1 - c_{13}^2 s_{23}^2 ) c_{34}^2 \sin^22\theta_{24} \,  
\\ & - \left.
c_{13}^2 c_{23}\sin 2\theta_{23} \sin 2\theta_{24} \sin 2 \theta_{34}\left( \frac{1}{2}-c_{13}^2 s_{23}^2  \right) \cos\delta_{24}
 \right] \sin^2\Delta_{31}  \\ 
 & - \frac{1}{4} c^2_{13} c_{24} \sin 2\theta_{23}  \sin 2\theta_{24} \sin 2\theta_{34} \sin\delta_{24} \sin 2 \Delta_{31}.
\end{split}
\label{eq:average}
\end{equation}
\end{enumerate}

\begin{figure}[t!]
\includegraphics[width=0.32\textwidth]{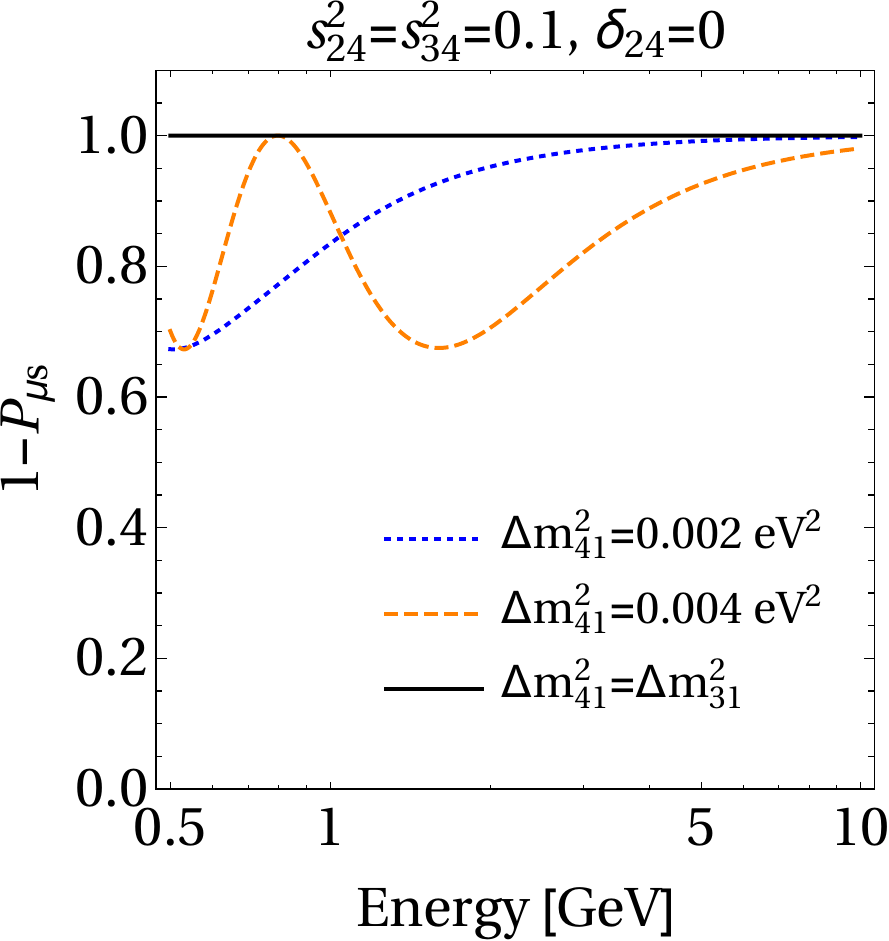}
\includegraphics[width=0.32\textwidth]{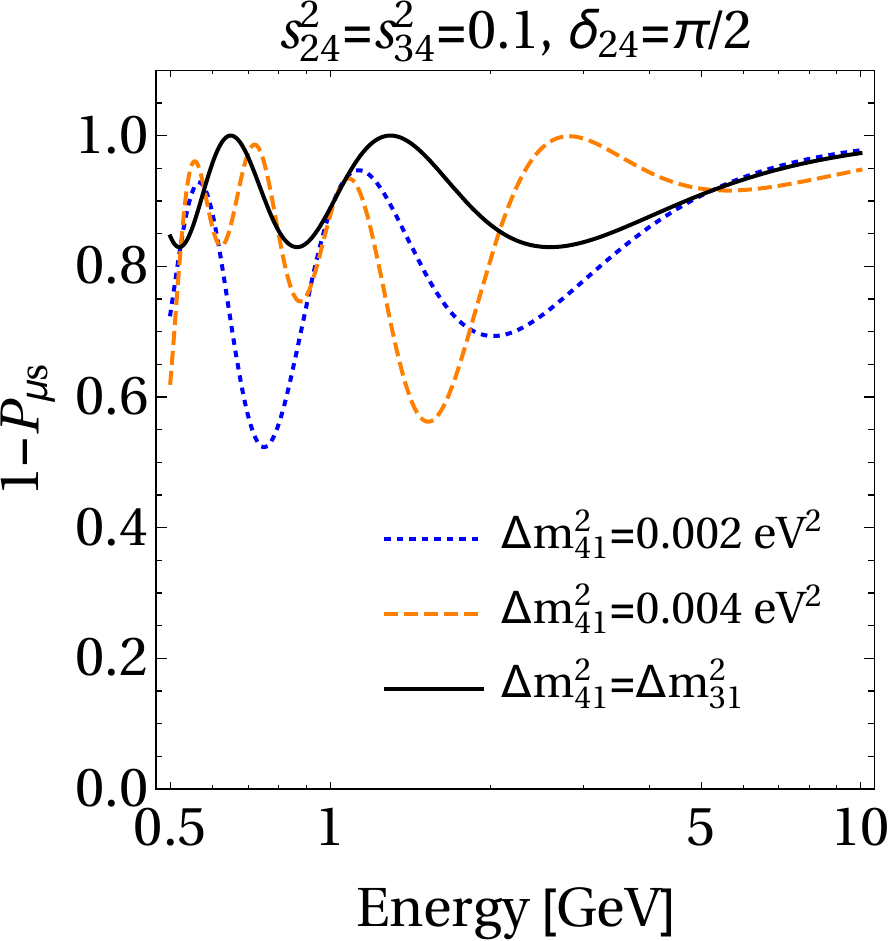}
\includegraphics[width=0.32\textwidth]{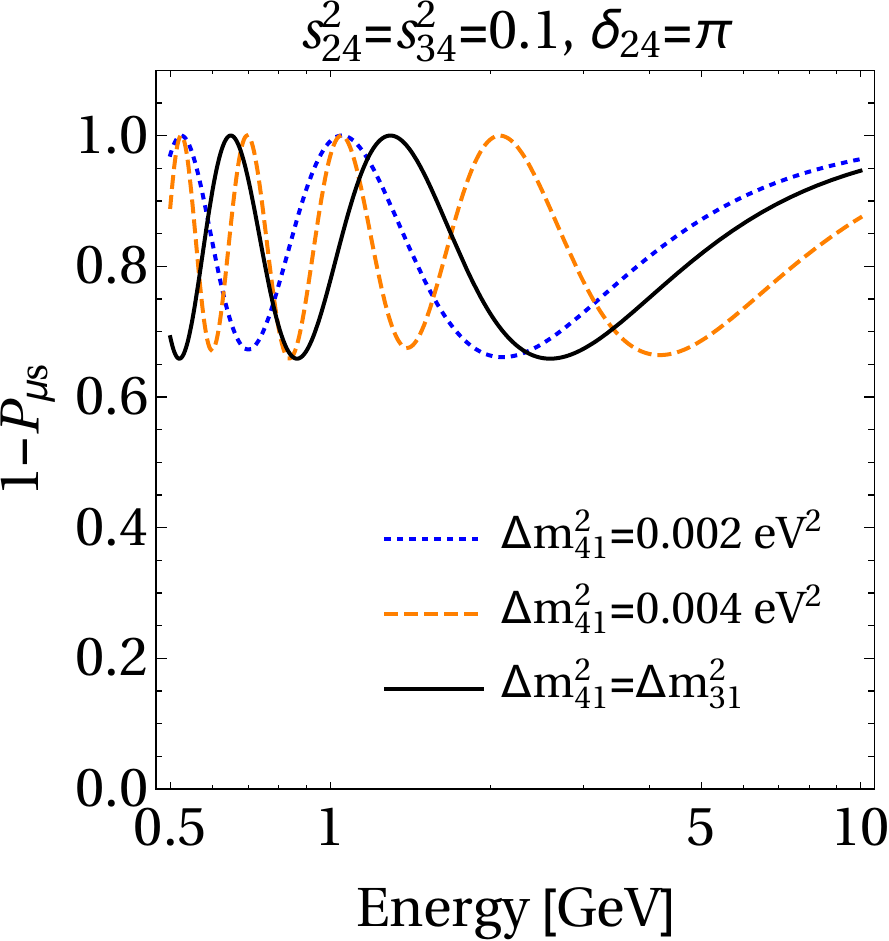}
\caption{\label{fig:prob-plot} Oscillation probability in the $\nu_\mu \to \nu_s$ channel, in vacuum. The different panels correspond to different values of the new CP-violating phase $\delta_{24}$, while the different lines shown in each panel correspond to different values of the active-sterile mass splitting $\Delta m^2_{41}$, as indicated in the legend. The rest of the oscillation parameters have been fixed to: 
$\Delta m^2_{31} = 2.48\times 10^{-3}\; \textrm{eV}^2 \,;~ 
\sin^2{\theta_{23}} = 0.5\,;~ 
\sin^2{2\theta_{13}} = 0.084\, ;$ and $\sin^2{\theta_{24}}  = \sin^2{\theta_{34}} = 0.1$. 
}
\end{figure}

As mentioned above, a destructive interference between the standard and non-standard contributions to the oscillation amplitude is possible for certain values of the active-sterile mixing parameters and, in particular, for certain values of the CP phase $\delta_{24}$. This is shown in Fig.~\ref{fig:prob-plot} for different values of $\Delta m^2_{41}$ around the atmospheric scale, when the oscillation probability simplifies to Eq.~(\ref{eq:dm41todm31}). The solid lines in all panels have been obtained for $\Delta m^2_{41}=\Delta m^2_{31}$: notice that a cancellation of the oscillation amplitude takes place in this case for $\dnew = 0$, as shown in the left panel in Fig.~\ref{fig:prob-plot}. In this case, the contribution from the interference (last term in Eq.~\ref{eq:dm41todm31}) is negative and cancels almost exactly the two other contributions to the oscillation probability. In fact, it is straightforward to show that, in the limit $c_{13}=c_{24}=c_{34} =1$, the amplitude of the oscillation is proportional to $c^2_{23}|s_{24}c_{23}-s_{34}s_{23}e^{i\dnew}|^2$, which vanishes exactly if $\dnew=0$ and $s_{24}c_{23}=s_{34}s_{23}$. This cancellation is only partial (or negligible) for other values of the CP phase, as expected, and this can be seen from the middle and right panels in the figure. For other values of the active-sterile mass splitting the oscillation pattern is more complex, as shown by the dotted blue and dashed yellow lines in Fig.~\ref{fig:prob-plot}. In the most general case, the dependence of the probability with the energy becomes non-trivial due to the interference of different terms oscillating at different frequencies. Moreover, as we will see in Sec.~\ref{sec:results} the cancellation in the probability can also be severe in the limit $\Delta m^2_{41} \ll \Delta m^2_{31}$.

Given the strong limits that have been set on the $\theta_{24}$ angle by the oscillation experiments looking for oscillations involving a sterile neutrino in the eV scale, it is worth to address explicitly the case when $\theta_{24}\to 0$. Under this assumption, the probability simplifies considerably with respect to the expression in Eq.~(\ref{eq:prob-final}):
\begin{equation}
P_{\mu s} (\theta_{24}\to 0) = 
c_{13}^4\sin^2 2\theta_{23} s^2_{34}\sin^2\Delta_{31}.
\label{eq:th24to0}
\end{equation}
In contrast with Eq.~(\ref{eq:prob-final}), in this case there is no sensitivity to $\delta_{24}$ and, most importantly, there is no dependence with the sterile mass-squared splitting. The oscillations in this case are solely driven by the atmospheric mass-squared splitting, and the size of the effect is directly proportional to $s_{34}^2$. Moreover, the dependence with the standard oscillation parameters goes as $c^4_{13} \sin^22\theta_{23}  \sim \mathcal{O}(1)$.

Finally, it is worth to mention that matter effects will modify the oscillation probability in Eq.~(\ref{eq:prob-final}). We have checked that the size of these modifications is relatively small and, therefore, the vacuum probabilities are precise enough to understand the behaviour of the numerical simulations in the following sections. However, in our numerical analysis, matter effects have been properly included using a constant matter density  of $2.96~\textrm{g}\cdot\textrm{cm}^{-3}.$

%
\section{Simulation}
\label{sec:simulation}

In contrast to usual analyses searching for signals of sterile neutrino oscillations at short distances, in this work we want to take advantage of the capabilities of the DUNE far detector, located at a distance of $L=1300$~km from the source. In particular, we focus on the potential of NC measurements to discriminate between the $3$-flavor and $4$-flavor scenarios. To this end, we rely on the excellent capabilities of the DUNE far detector to discriminate between charged-current (CC) and NC events. All the simulations in the current work have been performed using a modified version of the GLoBES~ \cite{Huber:2004ka,Huber:2007ji} library which includes a new implementation of systematic errors as described in Ref.~\cite{Coloma:2012ji}. The neutrino oscillation probabilities in a 3+1 scenario have been implemented using the new physics engine available from Ref.~\cite{globes-web}.

In our simulation of the signal, we have computed separately the contributions to the total number of events coming from $\nu_e$, $\nu_\mu$ and $\nu_\tau$ NC interactions at the detector. For simplicity, we have assumed a 90\% flat efficiency as a function of the reconstructed visible energy. The experimental observable for a NC event is a hadronic shower with a certain visible energy (energy deposited in the detector in the form of a track and scintillation light). The correspondence between a given incident neutrino energy and the amount of visible energy deposited in the detector has to be obtained from the simulation of neutrino interactions and detector reconstruction of the particles produced in the final state. To this end, we use the migration matrices provided by the authors of Ref.~\cite{DeRomeri:2016qwo}, which were obtained using the LArSoft simulation software~\cite{Church:2013hea}. The authors of Ref.~\cite{DeRomeri:2016qwo} used bins in visible energy of 50~MeV for the reconstructed energy of the hadron shower, as opposed to the DUNE CDR studies where wider bins of 125~MeV were considered~\cite{Acciarri:2015uup}. In the present work we have considered two sets of matrices: the original set provided by the authors of Ref.~\cite{DeRomeri:2016qwo}, with 50~MeV bins, and a (more conservative) rebinned version of these matrices where the bin size was increased to 250~MeV. We performed our simulations for the two options (with 50 MeV bins and 250 MeV bins) and found similar results for the two sets of matrices. Therefore, in the following we will adopt the more conservative 250~MeV bin size as our default configuration.

The main backgrounds for this search would be $\nu_e$, $\nu_\mu$ and $\nu_\tau$ CC events that might be mis-identified as NC events. We have assumed that the background rejection efficiency for CC events is at the level of $90\%$. However, this is probably a conservative estimate: for instance, muons leave long tracks in liquid Argon (LAr) that are difficult to misidentify as NC events, except when they have very low energies or are not completely contained in the detector. On the other hand, the active neutrino flavors would be affected by standard oscillations. Consequenly, the number of $\nu_\mu$ CC events would be largely suppressed since most of the initial muon neutrinos have oscillated to tau neutrinos by the time they reach the detector. Given the energetic neutrino flux at DUNE, some of the oscillated $\nu_\tau$ flux will interact at the detector via CC, producing $\tau$ leptons. In most of the cases ($ \approx 65\%$), the $\tau$ decays hadronically producing a shower: these events constitute an irreducible background and consequently no rejection efficiency has been assumed in this case. We have assumed a Gaussian energy resolution function for the $\nu_\mu$ and $\nu_e$ background contributions, following the values derived in Ref.~\cite{DeRomeri:2016qwo} from LArSoft simulations, while the hadronic showers produced from hadronic tau decays have been smeared using the same migration matrices as for the NC signal.

\begin{table}[ht!]
\setlength{\tabcolsep}{7pt}
\renewcommand\arraystretch{1.5}
\begin{tabular}{ c  c  ccc  }
\hline\hline
& Signal & \multicolumn{3}{c }{Background} \cr 
& $(N^{\nu_e}_{NC} + N^{\nu_\mu}_{NC} + N^{\nu_\tau}_{NC})$ & 
$N^{\nu_e}_{CC}$ & $N^{\nu_\mu}_{CC}$  & $N^{\nu_\tau}_{CC}$ \\ \hline 
$\nu$ mode & 6489 & 129 &  751 & 140 \\ 
$\bar\nu$ mode & 2901 & 22 &  301 &  39 \\ \hline\hline
\end{tabular}
\caption{\label{tab:events} Expected total number of events with a (reconstructed) visible energy between 0.5 and 8 GeV at the DUNE far detector. The number of events is shown for the signal and background contributions separately. This corresponds to 7~yrs of data taking (equally split between neutrino and antineutrino running modes) with a 40~kton detector and 1.07~MW beam power, yielding a total of of $300$~kt$\cdot$MW$\cdot$yr. In all cases, signal and background rejection efficiencies have already been accounted for. In the case of $ N^{\nu_\tau}_{CC}$, the number of events already includes the branching ratio for hadronic $\tau$ decays. Usual oscillations (in the three-family scenario) have been considered in the computation of the backgrounds, setting $\theta_{23}=42^\circ$ and the rest of the oscillation parameters in agreement with their current best-fit values.  }
\end{table}

The expected total number of signal and background events is summarized in Table~\ref{tab:events}, where the different background contributions are shown separately for clarity. As can be seen from this table, the largest background contribution comes from $\nu_\mu$ CC events mis-identified as NC, due to the large flux available at the far detector, while the contributions coming from $\nu_e$ and $\nu_\tau$ CC events are much smaller and approximately of equal size. In all cases, both signal and backgrounds receive contributions from right- and wrong-sign neutrino events due to the intrinsic contamination of the beam. The number of events has been computed for visible energies between 0.5~GeV and 8~GeV, which is the region used in our analysis, using the beam configuration with 80~GeV protons as in Ref.~\cite{Alion:2016uaj}. Additional experimental details for the DUNE setup considered in this work can be found in Refs.~\cite{Acciarri:2015uup,Alion:2016uaj}. 
%
\begin{figure}[t!]
\includegraphics[width=0.5\textwidth]{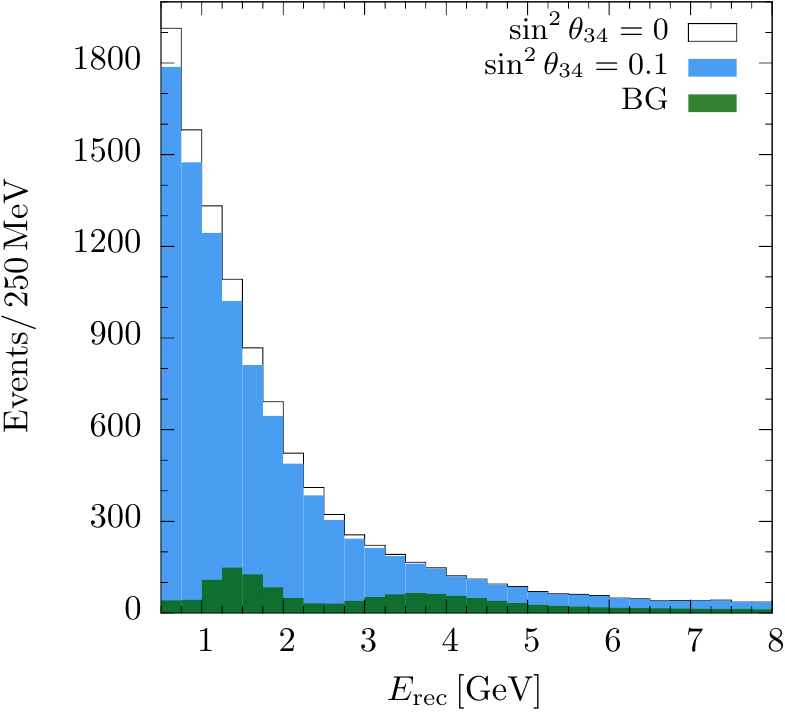}
\caption{\label{fig:event-plot} Expected signal and background event distributions, as a function 
of the reconstructed visible energy, after efficiencies and detector reconstruction. The white 
histogram shows the expected number of NC events in the $3$-family standard scenario, while the 
blue (light gray) histogram shows the expected number of NC events for  $\sin^2{\theta_{34}}=0.1$, 
$\theta_{14} = \theta_{24} = 0$. The expected distribution for background events (CC mis-identified as NC) 
is given by the green (dark gray) histogram. }
\end{figure}

The expected NC event distributions are shown in Fig.~\ref{fig:event-plot}, as a function of 
the (reconstructed) visible energy, for the three-family scenario (white histogram) and for the case when there 
is a sizable mixing angle with the sterile neutrino (blue/light gray histogram). As expected, a 
depletion in the number of events can be observed in the $3+1$ case with respect to the 
three-family scenario. Moreover, the events pile up at low energies due to the energy carried away 
by the outgoing neutrino in the final state. One can also see that the energy distribution of the background (shown by the green/dark gray histogram) 
is dictated by the standard oscillations suffered by the active neutrinos as they propagate to the 
far detector, which is well-known. In this case, all particles in the final state would be observed, 
and there is practically no pile-up at low energies. Due to this, the sensitivity to oscillations in 
the $\nu_\mu \to \nu_s$ channel is enhanced when some energy information is included in the fit, as 
we will see in the next section. This is exploited in our numerical analysis implementing a binned 
$\chi^2$ in the visible (deposited) energy in the detector. 

The effect of systematic uncertainties is accounted for through the addition of pull-terms to the 
$\chi^2$. In addition to an overall normalization uncertainty for the signal and background (which 
is bin-to-bin correlated), a shape uncertainty for the signal (bin-to-bin uncorrelated) has been 
included to account for possible systematic uncertainties related to the shape of the event 
distributions. Moreover, all nuisance parameters are taken to be uncorrelated between the neutrino 
and antineutrino channels as well as between the different contributions to the signal and/or 
background events. Unless otherwise stated, the final $\chi^2$ is obtained after marginalization 
over the nuisance parameters and the relevant standard oscillation parameters 
($\sin^22\theta_{23},\sin^22\theta_{13},\Delta m^2_{31}$) within current 
experimental uncertainties~\cite{Esteban:2016qun,Capozzi:2016rtj,Forero:2014bxa}. 
Specifically, we consider the following Gaussian priors: $ 
\sigma(\sin^22\theta_{13}) = 0.005$, $\sigma(\sin^22\theta_{23}) =  0.05$ and $\sigma(\Delta 
m^2_{31})/\Delta m^2_{31} = 0.04$. Unless otherwise specified, we have assumed a conservative $10\%$ 
gaussian prior for all nuisance parameters, included as pull-terms in the $\chi^2$. In practice, 
however, the cancellation of systematic errors in the NC channels is expected to be extremely 
efficient, since the near detector can be used to measure the \emph{same convolution} of the flux 
and cross section as in the far detector. This contrasts with oscillation measurements in appearance 
mode ($\nu_\alpha \to \nu_\beta$) using CC data, where the initial and final neutrino flux spectrum 
(and flavor) differ due to the impact of standard oscillations, making the cancellation of 
systematic uncertainties extremely challenging\footnote{For a recent review of the challenges that 
long-baseline experiments have to meet regarding systematic uncertainties see 
Ref.~\cite{Alvarez-Ruso:2017oui}. }. In spite of these difficulties, the DUNE collaboration expects to reach a 
precision at the percent level in the $\nu_\mu \to \nu_e$ and $\bar\nu_\mu \to \bar\nu_e $ 
appearance channels. In view of this, we expect the 5\% - 10\% values considered in this work for 
the NC sample to be conservative. 

Before concluding this section, let us comment on the relevance of the near detector data and its possible impact on the fit. In this work, we have not simulated the near detector explicitly: its design is still undecided and its expected performance is therefore unclear yet. A detailed simulation of the near-far detector data combination is beyond the scope of this work and can ultimately be performed only by the experimental collaboration. In this work, instead, we have assumed that the oscillations due to the new state have not developed yet at the near detector. For neutrino energies in the region around 2-3~GeV, and for a near detector located at a distance of $L\sim \mathcal{O}(500)$~m, this is a valid approximation as long as $ \Delta m^2_{41} < 1$~eV$^2$. Under this assumption, the near detector measurements will provide a clean determination of the convolution of the NC cross section and the muon neutrino flux, which can then be extrapolated to the far detector with a small uncertainty. At this point, it should be mentioned that our assumed prior uncertainties for the systematic errors in the fit would correspond to the values used for the analysis of the far detector event rates. Thus, they correspond to estimates on the size of the final systematic errors that have to be propagated to the far detector, once the near detector data has already been accounted for. Finally, it should also be stressed that in the case that $\theta_{14} = \theta_{24} = 0$ there would be no effect on the near detector data regardless of the new mass-squared splitting. The reason is that, as it was shown in Eq.~\eqref{eq:th24to0}, the dependence with $\Delta m^2_{41}$ drops from the oscillation probabilities: this guarantees no effect at the near detector, while at the far detector data the oscillation would be driven by the atmospheric scale. Thus, in this case the effect in the oscillation would be observable for large enough $\theta_{34}$.

%
\section{Results}
\label{sec:results}
In this section we show our numerical results for the expected sensitivities to the new mixing parameters in the different scenarios discussed  in Sec.~\ref{sec:probabilities}. By the time DUNE starts taking data the constraints on the sterile mixing angles $\theta_{14}$ and $\theta_{24}$ might be very tight. Nevertheless DUNE is also sensitive to the $\theta_{34}$ sterile mixing angle, which is currently the less constrained among the three sterile-active mixing angles. Therefore, we initially consider the simpler case where two of the new mixing angles fixed to zero, $\theta_{14}=\theta_{24}=0$ and study the sensitivity of the DUNE experiment to $\theta_{34}$. Next we proceed to turn on the mixing angle $\theta_{24}$ and determine for which values of $\theta_{24} - \Delta m^2_{41}$ the three-family hypothesis could be rejected. We finalize this section by showing the expected limits that could be derived simultaneously on the two mixing angles $\theta_{24}$ and $\theta_{34}$, for different values of the active-sterile mass-squared splitting.

\subsection{Sensitivity to $\theta_{34}$, for $\theta_{24} = 0$}
\label{sec:results-1}

Under the assumption $\theta_{24}=\theta_{14}=0$, the expression for the vacuum sterile neutrino appearance probability is given by Eq.~(\ref{eq:th24to0}) and does not depend on any of the new oscillation frequencies induced by the sterile, nor any of the CP-violating phases. An interesting question to ask in this case is if DUNE will be able to improve over current constraints on $\theta_{34}$, assuming that the experiment will measure event distributions in agreement with the expectation in the three-family scenario. In this case, the ``observed'' event distributions are simulated setting all $\theta_{i4}=0$, and are then fitted using increasing values of $\theta_{34}$. 

%
\begin{figure}[t!]
\includegraphics[width=0.5\textwidth]{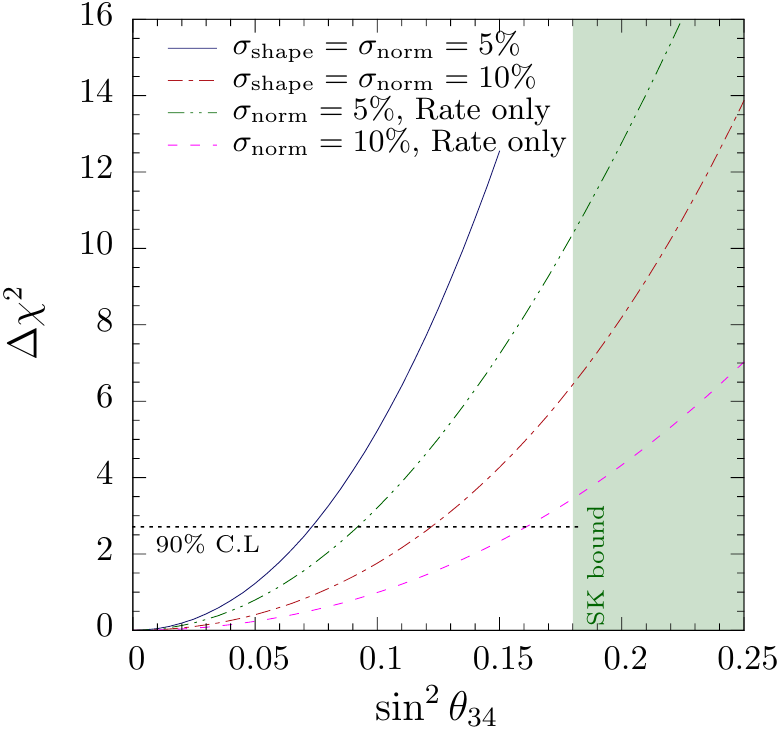}
\caption{\label{fig:theta34} Expected sensitivity to $\theta_{34}$ under the assumption $\theta_{14}=\theta_{24}=0$. The different lines correspond to different assumptions of systematical uncertainties, see text for details. The shaded region is disfavored at 90\% C.L. from Super-Kamiokande atmospheric data~\cite{Abe:2014gda}, $| U_{\tau 4} |^2< 0.18$ (at 90\% CL), which for $\theta_{14}, \theta_{24} = 0$ translates into the constraint $ \sin^2\theta_{34}<0.18$. The horizontal dotted line indicates the value of the $\Delta \chi^2$ corresponding to 90\% C.L. for 1 d.o.f..}
\end{figure}

The sensitivity to $\theta_{34}$ is shown in Fig.~\ref{fig:theta34}. As seen in the figure, our results show a considerable dependence on the size and implementation of systematic errors. Assuming a (conservative) $10\%$ systematic error on both normalization ($\sigma_{norm}$) and shape ($\sigma_{shape}$), we find that DUNE will be sensitive down to values of $\sin^2{\theta_{34}} \sim 0.12$, at $90\%$ C.L. (1 d.o.f.). For comparison we also show the limit on this mixing angle obtained from atmospheric neutrino data collected by the Super-Kamiokande (SK) collaboration~\cite{Abe:2014gda}, for $\Delta m_{41}^2 > 0.1$ eV$^2$. If prior uncertainties could be reduced to the 5\% level for both normalization and shape errors, we find that DUNE would be able to improve over the SK constraint by more than a factor of two. It should be stressed that the DUNE constraint would be valid for any value of $\Delta m^2_{41}$, as long as $\theta_{24}, \theta_{14} \simeq 0$. In the next subsections we will study in detail the phenomenology in case $\theta_{24}\neq 0$.

Finally, the lines labeled as ``Rate only''  in Fig.~\ref{fig:theta34} do not include a binned $\chi^2$ and only consider the total event rates in the computation of the $\chi^2$. The change in sensitivity can be appreciated from the comparison between the dashed pink and dot-dashed red lines, for 10\% systematic errors (or between the dot-dot-dashed green and solid blue lines, for 5\% systematic errors). As can be seen, the inclusion of energy information leads to a noticeable improvement in the results. Therefore, in the rest of this section we will only consider a binned $\chi^2$, using equally-sized bins in visible energy, as described in Sec.~\ref{sec:simulation}.

\subsection{Rejection power for the three-family hypothesis, for $\theta_{24}, \theta_{34} \neq 0$ }
\label{sec:results-2}

\begin{figure}[t!]
\includegraphics[width=0.325\textwidth]{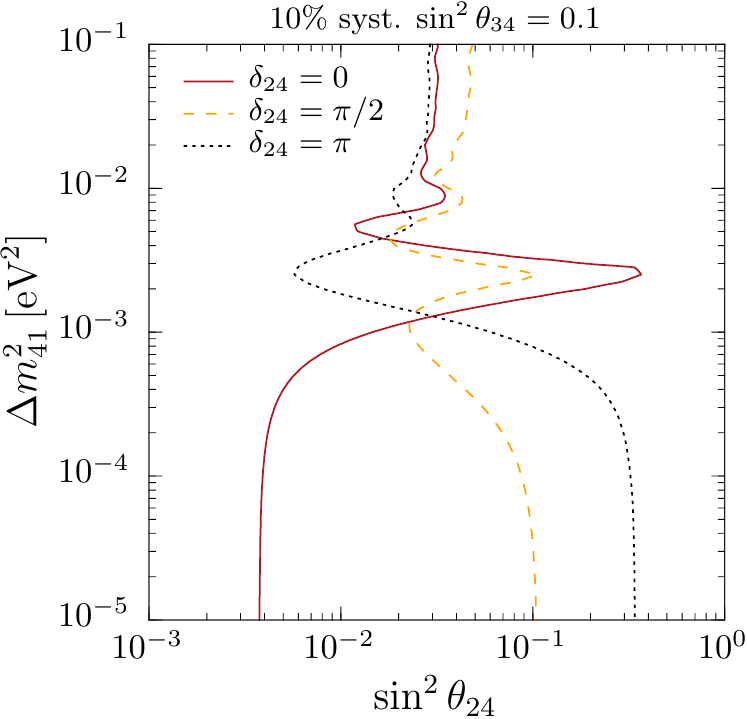}
\includegraphics[width=0.325\textwidth]{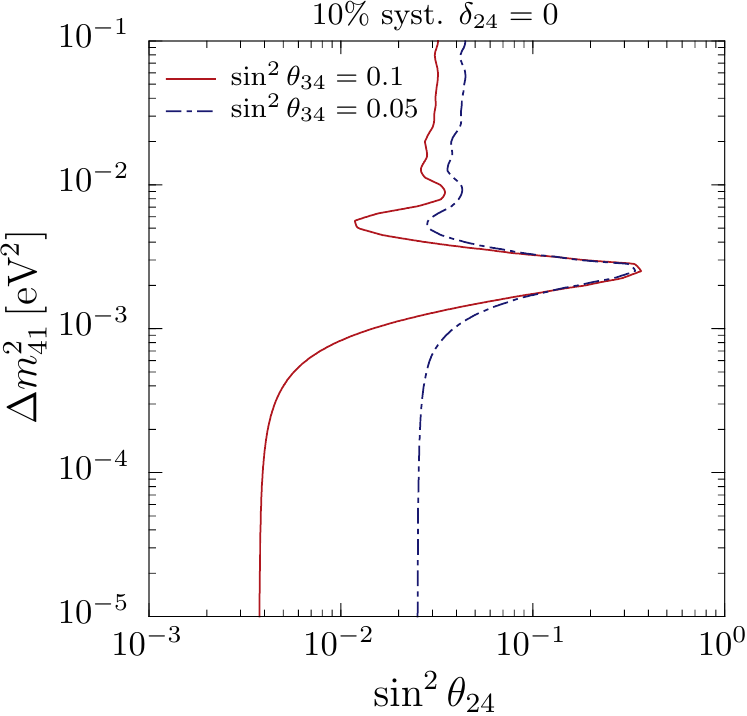}
\includegraphics[width=0.325\textwidth]{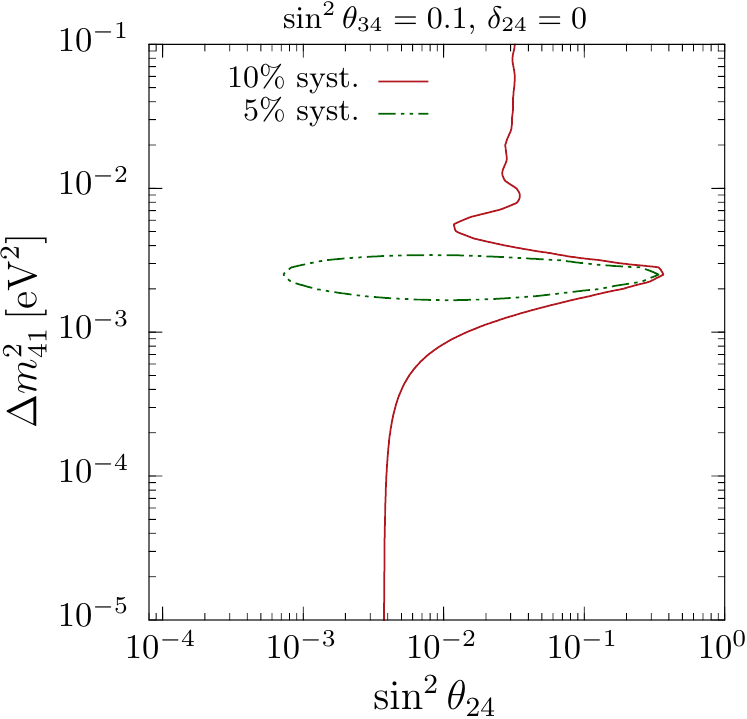}
\caption{\label{fig:analysis2} Rejection power for the three-family hypothesis, as a function of the assumed true values of $\Delta m^2_{41}$ and $\sin^2{\theta_{24}}$. The true value of $\theta_{34}$ has been set to a non-zero value in all cases, as indicated in the labels, while $\theta_{14}=0$ for simplicity. The contours indicate the sets of true values ($\theta_{24}$, $\Delta m^2_{41} $) for which the three-family hypothesis would be successfully rejected at 90\% C.L.. Left panel: dependence of the results with the true value of $\delta_{24}$. Central panel: dependence of the results with the true value of $\sin^22\theta_{34}$. Right panel: dependence of the results with the assumed priors for the systematic uncertainties. }
\end{figure}

The scenario where $\theta_{24}\ne0$ leads to a more interesting phenomenology, since in this case the oscillation probability also depends on the active-sterile mass-squared splitting. In this case, assuming as our true hypothesis a 3+1 with nonzero $\theta_{34}$ and $\theta_{24}$, it is relevant to ask if the experiment would be able to reject the three-family hypothesis. This is shown in Fig.~\ref{fig:analysis2}, as a function of the possible true values of $\Delta m^2_{41}$ and $\sin^2{\theta_{24}}$. The true value of $\theta_{34}$ is set to be nonzero, while $\theta_{14}=0$ is assumed for simplicity. In all panels, the expected events distributions are computed using the indicated values as \emph{true} input values. The obtained ``observed'' event distributions are then compared to the expected result in the three-family scenario, \emph{i.e.}, in absence of a sterile neutrino. The contours indicate the sets of true values ($\theta_{24}$, $\Delta m^2_{41} $) for which the three-family hypothesis would be successfully rejected at 90\% C.L.. The different panels in Fig.~\ref{fig:analysis2} show the dependence of our results with respect to different parameters: the true value of $\delta_{24}$ (left panel), the true value of $\theta_{34}$ (central panel); and the assumed priors for the systematic uncertainties (right panel).

As explained in Sec.~\ref{sec:probabilities}, if both $\theta_{24}$ and $\theta_{34}$ are different from zero, the oscillation probability $P_{\mu s}$ also depends on the value of the CP phase $\delta_{24}$. Such dependence can be appreciated by comparing the three lines shown in the left panel in Fig.~\ref{fig:analysis2}, corresponding to different true values of $\delta_{24}$. The same true value of $\theta_{34}$ and the same implementation of systematic uncertainties have been assumed for all lines (indicated by the top label). As shown in Fig.~\ref{fig:prob-plot} (see also Eq.~\eqref{eq:dm41todm31}), for values of $\Delta m^2_{41}\lesssim \Delta m^2_{31}$ there can be a large interference between the different contributions to the oscillation amplitude, depending on the value of $\delta_{24}$. For values of $\Delta m^2_{41}\simeq \Delta m^2_{31}$, this leads to a decreased sensitivity in this region of the parameter space for $\delta_{24} = 0$ with respect to the results obtained for $\delta_{24} = \pi$. The interference has the opposite effect in the region $\Delta m^2_{41}\ll \Delta m^2_{31}$: for negative values of $\cos\delta_{24}$ the second term in Eq.~\eqref{eq:dm41to0} is negative and suppresses the probability, leading to worse results for $ \delta_{24} = \pi$. In fact, it can be easily shown that, in the limit $\theta_{23}=\pi/4$, $c_{13}^2 = 1$ and at the first oscillation maximum ($\sin^2\Delta_{31} = 1$) the oscillation probability in Eq.~\eqref{eq:dm41to0} approximates to
\begin{equation}
\label{eq:approx-smallDm}
   P_{\mu s}\approx  c^2_{24}(s^2_{34} + 2 s_{24} s_{34} c_{34} \cos \delta + s^2_{24} c^2_{34}) \ ,
\end{equation} 
where the effect of the interference term can be easily appreciated. 

Conversely, in the limit where the new frequency is averaged-out ($\Delta m^2_{41}\gg \Delta m^2_{31}$) the results show a very mild dependence with the value of $\delta_{24}$. This can be easily explained from the expression in Eq.~\eqref{eq:average}, which shows two terms that depend on the value of $\delta_{24}$: the first one is directly proportional to $(c_{13}^2 s_{23}^2 - 1/2) \simeq \mathcal{O}( \delta\theta_{23} - s_{13}^2/2 )$, where $\delta\theta_{23} \equiv \theta_{23} - \pi/4$, and is therefore very suppressed; while the second term is proportional to $\sin2\Delta_{31}$ and it is completely off-peak at the first oscillation maximum. In fact, in the same limit ($\theta_{23}=\pi/4$, $c_{13}^2 = 1$) and at the first oscillation maximum it is easy to show that the term proportional to $\cos \delta_{24}$ in the oscillation probability in Eq.~\eqref{eq:average} is additionally suppressed with $\cos 2 \theta_{23}$, which is small for $\theta_{23}$ near maximal mixing.

The central panel in Fig.~\ref{fig:analysis2} shows the dependence of the results with the true value of $\theta_{34}$. In this case, all priors for the systematic uncertainties are set at the 10\% and we have fixed $\delta_{24} = 0$. As shown in the figure, in the region where $\Delta m^2_{41} \ll \Delta m^2_{31}$ there is a strong dependence of the results with the true value of $\theta_{34}$, while the contours do not show large variations for larger mass splittings. This behaviour can again be easily traced back to the approximate oscillation probabilities in Sec.~\ref{sec:probabilities}.

Finally, the right panel in Fig.~\ref{fig:analysis2} shows the dependence of the results with the 
assumed priors for the systematic uncertainties. In this panel, the true values of $\delta_{24}$ and 
$\theta_{34}$ have been set as indicated in the top label. The solid line uses our default 
implementation for the systematic uncertainties, where all priors are set to 10\% for both the shape 
and normalization and for both signal and background. The dot-dashed line, on the other hand, shows 
the room for improvement if all prior uncertainties can be reduced down to 5\%. As can be seen from 
the figure, the improvement is dramatic and leads to a successful rejection of the three-family hypothesis in practically all the parameter space, with the sole exception of the region around $\Delta m^2_{41} \simeq \Delta m^2_{31} $ (which is very difficult to reject, since this is the region where significant cancellations can take place for $\delta_{24}=0$).

\subsection{Expected allowed regions in the $\theta_{24} - \theta_{34}$ parameter space}
\label{sec:results-3}

If the observed event distributions show an agreement with the three-family expectation, one would proceed to derive a limit on the mixing angles $\theta_{24}$ and $\theta_{34}$. However, as we saw in Sec.~\ref{sec:probabilities} the oscillation probabilities show a large dependence with the new CP-violating phase $\delta_{24}$, and strong cancellations between the different contributions may occur. The effect of the cancellations is much more severe in the limit $\Delta m^2_{41} \to 0$ than for larger values of the active-sterile mass splitting and, therefore, we expect very different results as a function of this parameter. 

\begin{figure}[t!]
\includegraphics[width=0.8\textwidth]{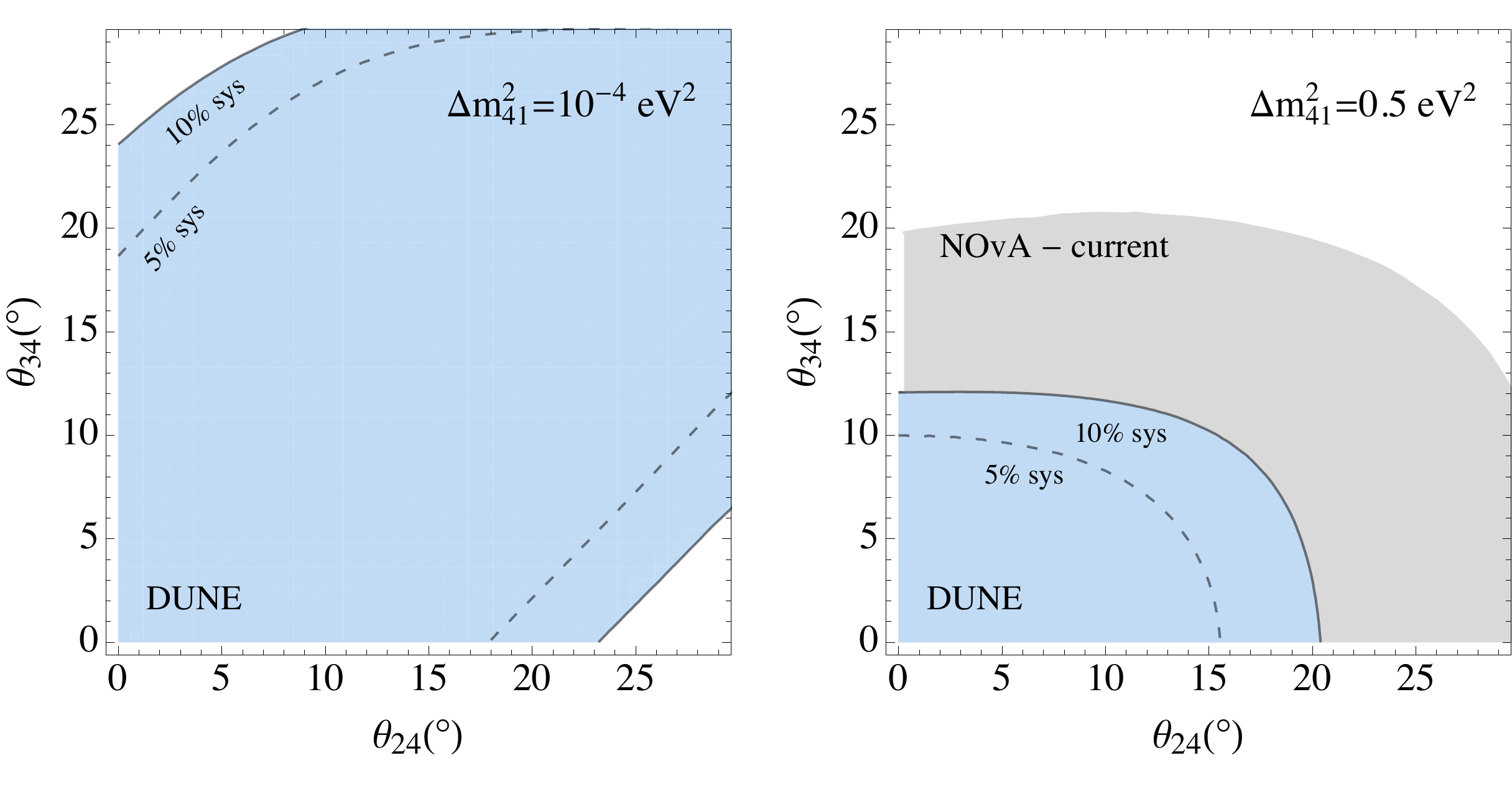}
\caption{\label{fig:th24th34} Expected sensitivity projected in the $\theta_{24} - \theta_{34}$ 
plane, for an active-sterile mass-squared splitting $\Delta m^2_{41} = 10^{-4}$~eV$^2$ (left panel) 
and for $\Delta m^2_{41} = 0.5$~eV$^2$ (right panel). The shaded regions correspond to the expected 
confidence regions allowed at 90\% C.L. (2 d.o.f.), for a simulation assuming $\theta_{i4}=0$ as 
true input values. The lines labeled as ``10\% sys'' (``5\% sys'') have been obtained assuming 10\% 
(5\%) prior uncertainties for the signal (both shape and normalization) and 10\% for the background 
(normalization only). For comparison, the right panel shows the latest results from the NOvA 
experiment from a NC search, also at the 90\% C.L.~\cite{Adamson:2017zcg}. }
\end{figure}

Figure~\ref{fig:th24th34} shows the expected allowed regions in the $\theta_{24}$ and $\theta_{34}$ plane if the observed event distributions are found to be in agreement with the three-family hypothesis. In this case, the ``observed'' event distributions are simulated assuming the three-family hypothesis, and fitted in a 3+1 scenario. The value of the $\chi^2$ function, for a given pair of test values $\theta_{24}-\theta_{34}$, is obtained after minimization over the new CP-violating phase $\delta_{24}$ and over all nuisance parameters. As for the mass splitting $ \Delta m^2_{41}$, it has been kept fixed during the fit to the test value indicated in each panel to show the difference in the results. For simplicity, we have also kept all the standard parameters fixed during the minimization procedure; however, minimization over the standard parameters is not expected to affect significantly the results shown here.

As shown in Fig.~\ref{fig:th24th34}, the resulting allowed regions are very different if the 
results are tested using $ \Delta m^2_{41} \ll \Delta m^2_{31}$ or a $\Delta m^2_{41}$ in the 
averaged-out regime. In the former case, a strong cancellation in the oscillation probability can 
always be achieved setting the value of $\delta_{24} \sim \pi$, as outlined in 
Sec.~\ref{sec:probabilities} and Sec.~\ref{sec:results-2}. Therefore, in this case it is not 
possible to disfavor large values of the new mixing angles. Only if the two mixing angles have very 
different values (\eg in the region $\theta_{24} \to 0, \theta_{34} \gtrsim 25^\circ$) the interference 
term would not be large enough to allow for an efficient cancellation in the probability. Thus, in this regime DUNE 
could disfavor just the upper left and lower right corner of the parameter space. Conversely, 
in the limit $\Delta m^2_{41} \gg \Delta m^2_{31}$ the impact of the new CP-violating phase 
$\delta_{24}$ is much milder and does not allow for a cancellation in the oscillation probability. A 
closed contour is therefore obtained in this case. For comparison, we show the currently allowed 
regions from an analysis of the NOvA far detector neutral-current data sample, taken from 
Ref.~\cite{Adamson:2017zcg}. As shown in the figure, DUNE is expected to improve over a factor of 
two with respect to the current allowed region set by NOvA. We also show two sets of lines for the 
DUNE experiment, which indicate the improvement in the results if the signal systematic uncertainties 
could be reduced below the 10\% level.

%
\section{Summary and conclusions}
\label{sec:conclusions}

The experimental anomalies independently reported in LSND, MiniBooNE, reactor and Gallium experiments have put the possible existence of an eV-scale sterile neutrino under intense scrutiny. In the near future a new generation of short-baseline experiments will come online to refute or confirm these hints, and will place strong constraints on the mixing of a light sterile neutrino with electron and muon neutrinos. Achieving similar bounds on the mixing with tau neutrinos is a much more difficult task, given the technical challenges associated to the production and detection of $\nu_\tau$. At long-baseline experiments, however, oscillations in the $\nu_\mu \to \nu_\tau$ channel guarantee that most of the beam will have oscillated into $\nu_\tau$ by the time it reaches the far detector, thanks to the atmospheric mass-squared splitting. By searching for a depletion in the number of neutral-current (NC) events measured at the far detector, experiments like NOvA or MINOS have been able to probe the mixing between $\nu_\tau$ and a fourth neutrino.

In this work, we have studied the potential of the future DUNE experiment to conduct a search for sterile neutrinos using the NC data expected at the \emph{far detector}, taking advantage of the excellent capabilities of liquid Argon to discriminate between charged-current and NC events. For simplicity, we have focused on a $3+1$ scenario, where only one extra sterile neutrino is introduced. In this case, the mixing matrix has to be extended including three additional mixing angles ($ \theta_{14}$, $\theta_{24}$ and $\theta_{34}$) and two CP-violating phases $\delta_{14}$ and $\delta_{24}$ (our parametrization is given by Eq.~\eqref{eq:paramet}). The oscillation probabilities will generally depend on an additional oscillation frequency dictated by the mass-squared splitting between the active and sterile states, $ \Delta m^2_{41}$. First, we have derived the oscillation probabilities in different regimes paying particular attention to the dependence with the new CP-violating phases. Unlike in other studies where the mass of the sterile was required to be at (or around) the eV scale, here we have allowed it to vary between $10^{-5}$~eV$^2$ and $10^{-1}$~eV$^2$; thus, in Eqs.~\eqref{eq:dm41to0}-\eqref{eq:average} we provide approximate expressions for the oscillation probabilities in three different regimes, depending on the mass of the sterile state: (i) $\Delta m^2_{41}\to 0 $; (ii) $\Delta m^2_{41}\simeq \Delta m^2_{31}$; and (iii) $\Delta m^2_{41}\gg \Delta m^2_{31}$. 

We have then proceeded to simulate the expected sensitivity of the DUNE experiment using the expected NC events collected at the far detector. We have studied the variation of our results with the implementation and size of the systematic errors. The details of our numerical simulations and the $\chi^2$ implementation can be found in Sec.~\ref{sec:simulation}. 

First, working under the assumption $\theta_{24} = \theta_{14} = 0$, we have determined the sensitivity of the DUNE experiment to the third mixing angle $\theta_{34}$. In this case, the oscillation probability is independent of the new CP-violating phases; furthermore, oscillations are solely driven by $\Delta m^2_{31}$, see Eq.~\eqref{eq:th24to0}. We find that DUNE will be able to improve over current constraints on this parameter set by the SK experiment, and will be sensitive to values of $\sin^2\theta_{34}\sim 0.12$ (at 90\% CL) for our default implementation of systematic uncertainties. If systematic errors could be reduced down to 5\%, the experimental sensitivity would reach $\sin^2\theta_{34} \sim 0.07$ (at 90\% CL).

Next we proceeded to study the case where $\theta_{24} \neq 0$. In this case, the oscillation probabilities depend on the active-sterile mass-squared splitting. The phenomenology becomes more complicated and, in particular, strong cancellations in the probability can take place for certain values of  $\dnew$ and $\Delta m^2_{41}$. First, we considered the 3+1 scenario as the true hypothesis, and determined for which values of the mixing parameters DUNE would be able to reject the three-family scenario. Our results are summarized in Fig.~\ref{fig:analysis2}, where we show the dependence of the sensitivity with the CP phase $\dnew$, the mixing angle $\theta_{34}$ and the size of the systematic errors. We found that the sensitivity of the experiment to the presence of a sterile neutrino, measured as its ability to reject the three-family scenario, depends heavily on the value of the CP phase. For example, for $\Delta m^2_{41} = 10^{-4}$~eV$^2$, $\sin^2\theta_{34} = 0.1$ and $\dnew = 0$, DUNE would be able to reject the three-family scenario for $\sin^2\theta_{24}\lesssim 4\times 10^{-3}$~eV$^2$; conversely, for $\dnew =\pi$ (and assuming the same value for $\theta_{34}$ and $\Delta m^2_{41}$), $\theta_{24}$ could be almost two orders of magnitude larger and the three-family scenario would not be rejected by the data. The behaviour of our  results can be easily understood in terms of the oscillation probabilities, as explained in detail in Sec.~\ref{sec:results-2}.

Finally, we considered the opposite situation, and assumed that the experiment will find a result that is in agreement with the three-family expectation. In this case, we determined the allowed confidence regions that would turn from the analysis of the simulated data. Our results are shown in Fig.~\ref{fig:th24th34}. The simulated data were tested using two very different values of the active-sterile mass-squared splitting. In the averaged-out regime ($\Delta m^2_{41} \gg \Delta m^2_{31}$), a closed contour is obtained; we find that DUNE would be able to improve over NOvA constraints in this place by a factor of two or more, depending on the size of the systematic errors assumed. Conversely, in the case of $\Delta m^2_{41} \ll \Delta m^2_{31}$ the experimental results would allow values of $\theta_{24}$ and $\theta_{34}$ to be as large as $30^\circ$. The reason is, again, the possibility of having a strong cancellation in the oscillation probability, which could lead to a non-observable effect in the event distributions even in presence of very large mixing angles.

The DUNE experiment has unprecedented discrimination between neutral-current and charged-current events for a long-baseline experiment: this will allow for a measurement or constraint of the $\nu_\tau$ fraction of a possible sterile neutrino(s).  Given the difficulties associated to the production and detection of $\nu_\tau$'s, measurement or limiting this fraction by other means is very challenging.  In this paper, we show that the DUNE experiment can provide an excellent constrain or discover a sterile neutrino that primarily mixes with only the $\nu_\tau$.


\acknowledgements We warmly thank Michel Sorel for providing us with 
the smearing matrices needed to simulate the liquid Argon detector reconstruction for 
neutral-current events. PC also thanks Enrique Fernandez-Martinez for useful discussions. 
DVF is thankful for the support of S\~ao Paulo Research Foundation (FAPESP) 
funding Grant No. 2014/19164-6 and 2017/01749-6., and also for the URA fellowship that allowed him 
to  visit  the  theory  department  at  Fermilab  where  this  project  started. DVF was also 
supported by the U.S. Department Of Energy under contracts DE-SC0013632 and DE-SC0009973. This work 
has received partial support from the European Union's Horizon 2020 research and innovation 
programme under the Marie Sklodowska-Curie grant agreement No.~674896. This manuscript has been 
authored by Fermi Research Alliance, LLC under Contract No. DE-AC02-07CH11359 with the U.S. 
Department of Energy, Office of Science, Office of High Energy Physics. The United States Government 
retains and the publisher, by accepting the article for publication, acknowledges that the United 
States Government retains a non-exclusive, paid-up, irrevocable, world-wide license to publish or 
reproduce the published form of this manuscript, or allow others to do so, for United States 
Government purposes.

\appendix

\section{Complete expressions for the relevant mixing matrix elements in our parametrization}
\label{app}
Starting from the parametrization in Eq.~(\ref{eq:paramet}), the mixing matrix elements needed for the calculation of the sterile appearance probability are given by:

\be{eq:elements}
\begin{split}
U_{\mu 4}& = e^{-i \delta_{24}} c_{14} s_{24} \, , \\
U_{s 4} &= c_{14} c_{24} c_{34} \, , \\
U_{\mu 3} &= c_{13} s_{23} c_{24}  - 
 e^{-i( \delta_\textrm{CP} - \delta_{14} + \delta_{24})} s_{13} s_{14} 
s_{24} \, , \\
U_{s 3} &= -e^{-i( \delta_\textrm{CP} - \delta_{14})} s_{13} c_{24} c_{34} 
s_{14} -  e^{i \delta_{24}} c_{13} s_{23} s_{24} c_{34}   -  c_{13} c_{23} s_{34}.
\end{split}
\ee
For $\theta_{14}=0$, and using Eq.~(\ref{eq:elements}), we find the following useful expressions:
\be{eq:a1}
\begin{split}
|U_{s 3}|^2 &=c_{13}^2 \left( c^2_{23} s^2_{34} + \frac{1}{2} \sin 2\theta_{23} 
s_{24} \sin 2\theta_{34} \cos \delta_{24}  + s^2_{23} s^2_{24}c^2_{34} \right) \, ,\\
8~U_{\mu 4}^* U_{s 4} U_{\mu 3} U_{s 3}^* &=- c^2_{13}  c_{24} \sin 2\theta_{23}  \sin 2\theta_{24} \sin 2\theta_{34}  
e^{i\delta_{24}} - 2c^2_{13} s^2_{23} c^2_{34} \sin^2 2\theta_{24} . 
\end{split}
\ee

\bibliography{sterile_nu_DUNE}

\end{document}